# When to Commute During the COVID-19 Pandemic and Beyond: Analysis of Traffic Crashes in Washington, D.C.

Joanne Choi[1], Sam Clark[1], Ranjan Jaiswal[1], Peter Kirk[1], Sachin Jayaraman[1], and Huthaifa I. Ashqar[1,2,*]

[1] Computer Science and Electrical Engineering Department University of Maryland Baltimore County at Maryland, USA (joannechoii@gmail.com)

[2] Arap American University, Jenin, Palestine (huthaifa.ashqar@aaup.edu)

[*] Corresponding Author

## Abstract

Many workers in cities across the world, who have been teleworking because of the COVID-19 pandemic, are expected to be back to their commutes. As this process is believed to be gradual and telecommuting is likely to remain an option for many workers, hybrid model and flexible schedules might become the norm in the future. This variable work schedules allows employees to commute outside of traditional rush hours. Moreover, many studies showed that commuters might be skeptical of using trains, buses, and carpools and could turn to personal vehicles to get to work, which might increase congestion and crashes in the roads. This study attempts to provide information on the safest time to commute to Washington, DC area analyzing historical traffic crash data before the COVID-19 pandemic. It also aims to advance our understanding of traffic crashes and other relating factors such as weather in the Washington, DC area. We created a model to predict crashes by time of the day, using a negative binomial regression after rejecting a Poisson regression, and additionally explored the validity of a Random Forest regression. Our main consideration for an eventual application of this study is to reduce crashes in Washington DC, using this tool that provides people with better options on when to commute and when to telework, if available.

The study also provides policymakers and researchers with real-world insights that decrease the number of traffic crashes to help achieve the goals of The Vision Zero Initiative adopted by the district.

## Introduction

Washington, District of Columbia (D.C.) is a major metropolitan area and the capital of the United States of America. The National Capital Region, colloquially known as the "DMV" (D.C., Maryland, Virginia) is known to be a very transient area, due to how many workers reside within the district and the two states. The National Capital Region Transportation Planning Board (2019) reported that nearly two-thirds of commuters drove alone within the Washington metropolitan area, as opposed to other means of commuting, such as Metrorail or teleworking. Additionally, the average American's commute time is 27 minutes, while D.C. commuters' average is 43 minutes (Berkon, 2020). The combination of high traffic volume and extensive commute times could possibly be linked to the frequency of vehicle crashes.

COVID-19 has reshaped the workforce and the commuting patterns. Before COVID-19, about 20% of the American workforce worked from home (Parker et al., 2021). Currently, 71% of the same workforce are now working from home and 54% would like to continue to work from home after restrictions are lifted (Parker et al., 2021). Due to the shift from commuting to telecommuting, we believe that many employers will be more flexible with their employees, allowing them to create their own commuting schedule rather than commute to work on a daily basis.

Given the massive number of daily commuters that D.C. received prior to COVID-19, we propose developing a general regression model to predict the likelihood of vehicle collision based on various risk factors. We will analyze crash data from within D.C. from 2016 to 2019 with weather data to determine which factors contribute most to crashes. Our analysis could be used to inform commuters on the likelihood of vehicle collision on any given day, possibly impacting the decision on whether to commute into the office. This model is intended to help reduce the number of commuters on days that are predicted to be more likely to have vehicle collisions.

## Literature Review

Road safety regarding traffic accidents has been an issue for as long as traffic has existed. Many studies have analyzed vehicle collisions to determine the risk factors and how they may contribute to the likelihood and severity of car crashes (Abdel-Aty et al., 2006; Ashqar et al., 2021; Preusser et al., 2020). Some of these risk factors include weather conditions and time of day. By determining the likelihood and the severity of the impact each factor has on car collisions, society can adapt and improve current conditions and policies to lessen the risk of car accidents.

Weather is a major risk factor that could lead to vehicle collisions. As of 2016, adverse weather conditions at the time of crash accounted for an annual average of 16% of all car crash deaths (Saha et al, 2016). Drivers were found to get into more car accidents during weather conditions involving precipitation than during extreme heat conditions (Liu et al. 2017). The accident probability was approximately 5 times higher in negative temperature than in positive temperatures (Becker et al., 2020). Precipitation conditions include rain and snow depending on the season. Of the two variables, snow was found to have more impact on the likelihood of car crashes than rainfall (El-Basyounny et al, 2014). The number of vehicle collisions was positively correlated to increase in snowfall intensity.

However, some studies have concluded that less traffic accidents occur in presence of precipitation because of the change in traffic variation (Theofilatos, 2016). During rainfall or snow, it is speculated that drivers drive more carefully during wet conditions and there are less motorcyclists or pedestrians interrupting the traffic. In the D.C. area, commuters drive through varying weather conditions throughout the year therefore understanding the impact of weather on car crashes will be critical to commuters.

Another key risk factor that could potentially impact car crashes is time of day. A linear regression analysis done in New York City concluded that 4pm is the most likely time for a driver to get into an accident in New York City (Hopping, 2019). 4pm is the start of New York City's rush hour and the start of the busiest and heavy traffic times. Similarly, another study based on traffic in Connecticut has predicted that most crashes occurred in the afternoon between 4pm and 5pm (Mondal et al., 2020). Based on these previous studies, afternoon rush hour times after work tend to be when car crashes occur the most. Car crashes that occur during commute from work to home were found to be one of the major causes of injury and deaths among drivers (Bener et al., 2017). The traffic congestion occurring at these hours could be positively related to an increase in frequency of car accidents (Retallack et al., 2020). Another possible explanation for mid-afternoon car accidents is post-lunch sleepiness, which led to slower reaction times for drivers (Hao et al., 2016). By determining the most dangerous time of day in D.C. traffic could help commuters adjust their work schedule to avoid times when car crashes are most likely to occur.

The Washington D.C. commuters must be prepared to drive in every weather condition as the season changes. Especially during typical rush hour times, when drivers are more prone to car accidents, drivers must be more alert and careful during the commute. Understanding the impact of these risk factors on driving conditions and likelihood of car crashes will allow drivers in D.C. to make informed decisions on when or how they will commute to work. Additionally, evaluating these risk factors will help policy makers implement safety measures to reduce possibility of car accidents as well as severity of accidents.

## Dataset

For our analysis car crash data was obtained from Open Data DC's "Crashes in DC" (hereafter, "Crash Data"), a free online repository of data gathered by the Washington, District of Columbia government. The Crash Data is maintained by the District Department of Transportation (DDOT). The DDOT processes crash records reported by the Metropolitan Police Department's (MDP) crash data management system nightly. Each record indicates an individual crash reported by the MDP, and for the purpose of our analysis we considered each record as one crash. The Crash Data spans from 2016 to 2019. The Crash Data contains 245,136 records with no duplicate record found. We utilized the "REPORTDATE" column for our analysis, with the derived variables of Hour, Day, and Year. Figure 1 shows histograms demonstrating the distribution of the Hour, Day, and Month variables.

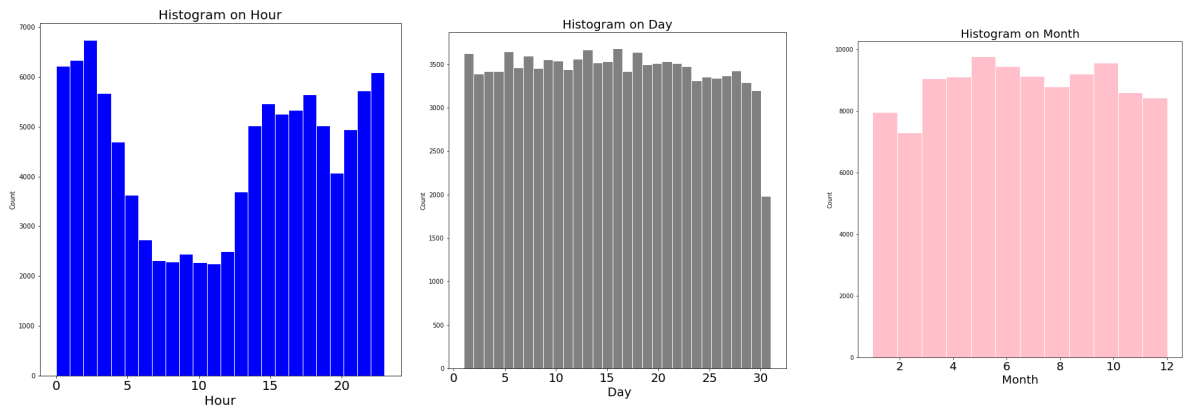

*Figure 1 Histograms of car crash distributions by Hour, Day, and Month*

For the analysis, daily weather data was collected from the Visual Crossing website for the years between 2016 and 2019. In the initial dataset there were some duplicates, created on the end date of Daylight saving for each year. The duplicated records with minimum values were removed to match with crash data. There were 1461 unique records after the data was cleansed. There were 17 columns including maximum, minimum, and daily average temperature (in Fahrenheit); wind speed and wind gust (in mph); precipitation, snow, and snow depth. Various studies show different weather parameters have significant impact on road accidents, but precipitation is the most important one (Elhenawy et al., 2021; Zou et al., 2021). Thus, for our study we only utilized "Precipitation", "Weather Conditions" along with the "Datetime" column. Note, data used here is daily average and does not indicate instant weather conditions. Figure 2 shows a sample of weather data, while Figure 3 shows visualizations of weather data by weather conditions and monthly precipitation.

| Date time | Maximum Temperature | Minimum Temperature | Temperature | Wind Chill | Heat Index | Precipitation | Snow | Snow Depth | Wind Speed | Wind Direction | Wind Gust | Visibility | Cloud Cover | Relative Humidity | Conditions |
|---|---|---|---|---|---|---|---|---|---|---|---|---|---|---|---|
| 2015-01-01 | 45.8 | 26.5 | 36.1 | 19.1 | 0.0 | 0.00 | 0.00 | 0.00 | 14.5 | 211.75 | 29.5 | 9.9 | 39.2 | 45.25 | Partially cloudy |
| 2015-01-02 | 48.5 | 36.3 | 41.0 | 32.8 | 0.0 | 0.00 | 0.00 | 0.00 | 9.0 | 249.29 | 18.3 | 9.9 | 81.7 | 52.35 | Overcast |
| 2015-01-03 | 41.9 | 34.2 | 39.0 | 30.7 | 0.0 | 0.43 | 0.00 | 0.00 | 6.4 | 64.79 | 0.0 | 6.2 | 94.6 | 77.60 | Rain, Overcast |
| 2015-01-04 | 65.8 | 41.9 | 53.1 | 36.1 | 0.0 | 0.21 | 0.00 | 0.00 | 20.5 | 216.38 | 32.3 | 9.0 | 96.3 | 77.52 | Rain, Overcast |
| 2015-01-05 | 51.8 | 29.7 | 40.0 | 19.7 | 0.0 | 0.00 | 0.74 | 0.74 | 22.3 | 309.54 | 35.2 | 9.9 | 62.9 | 33.94 | Snow, Partially cloudy |

*Figure 2 Sample of weather data*

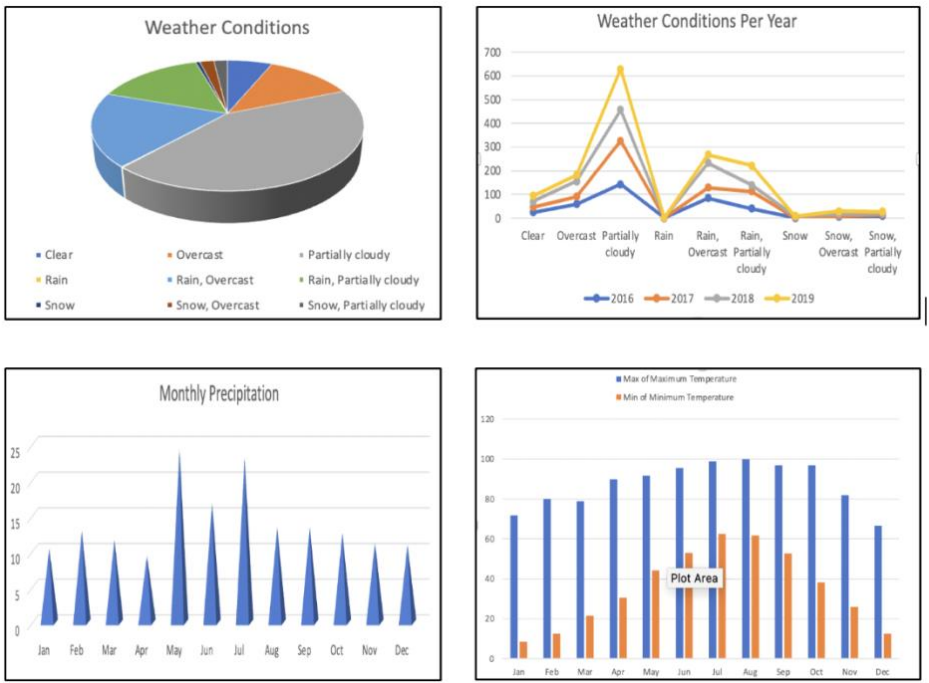

*Figure 3 Visualizations of weather data by weather conditions and monthly precipitation*

The Crash Data and Weather Data were merged on the date columns of each data set then grouped by Hour, Day and Month. The "CrashCount'' and "Precipitation indicator" variables were aggregated by sum giving the total number of crashes and the total number of days there was precipitation for the grouped rows. The Hour and Month variables were converted into dummy variables, respectively. The Day variable was converted into a weekday variable ranging from Monday to Sunday, and then converted into dummy variables. The Resulting data set contained 8,756 rows and 53 columns. Figure 4 shows a heatmap demonstrating the correlation of the final variables along with a histogram showing the distribution of the Crash Counts after grouping the data.

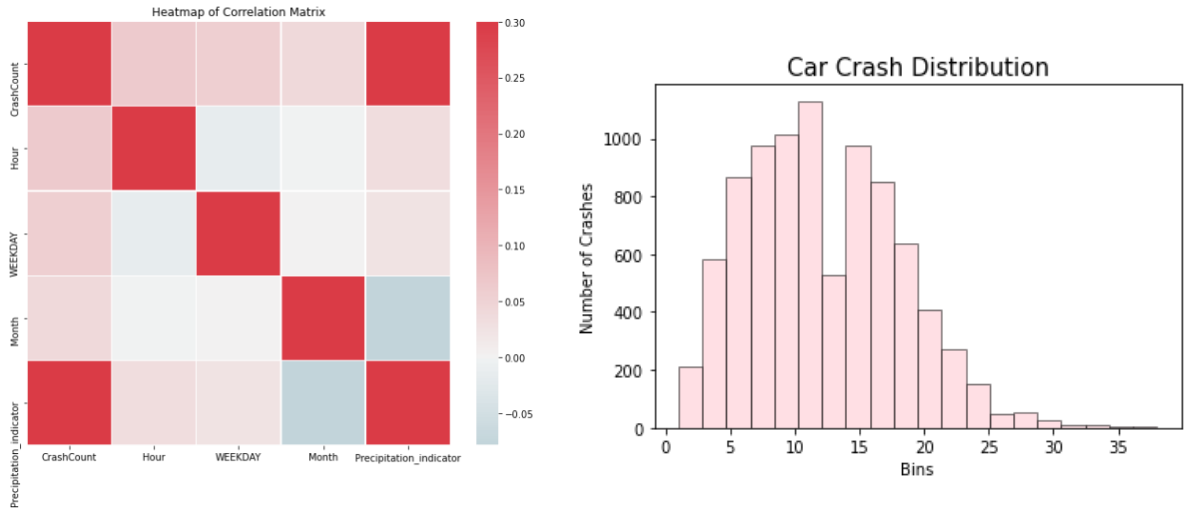

*Figure 4 Heatmap of final variables and bar chart of car crash distributions*

## Analysis and Results

### Random Forest

Random Forest (RF) Regression is one of the widely used machine learning algorithms, which involves constructing many trees from the training datasets. Prediction on RF Regression model is calculated from the average prediction across the decision trees. In the study, crash count is considered as a response variable whereas 69 other variables (precipitation, snow, 24 dummy variables for hour, 31 dummy variables for day and 12 dummy variables for month) are used as explanatory variables. Dataset was then split into 75% for training and 25% for testing the model. To get the best accuracy, we ran the model on a constant dataset for different values of 'n_estimators' i.e., number of trees used in the model and then observed the Mean absolute error and Accuracy.

Initially, the model experienced a steady decline in Mean Absolute error with increase in number of trees but after some point, a flat line was observed with maximum accuracy of ~60%. Then the model was trained on the training dataset with the best possible value of 'n_estimators' and predictions were made on the testing dataset. The scatter plot chart shows the actual daily crash count versus predicted crash count by the model (as shown in Figure 5). Based on the RF regression, precipitation was the most important feature in determining the car crashes.

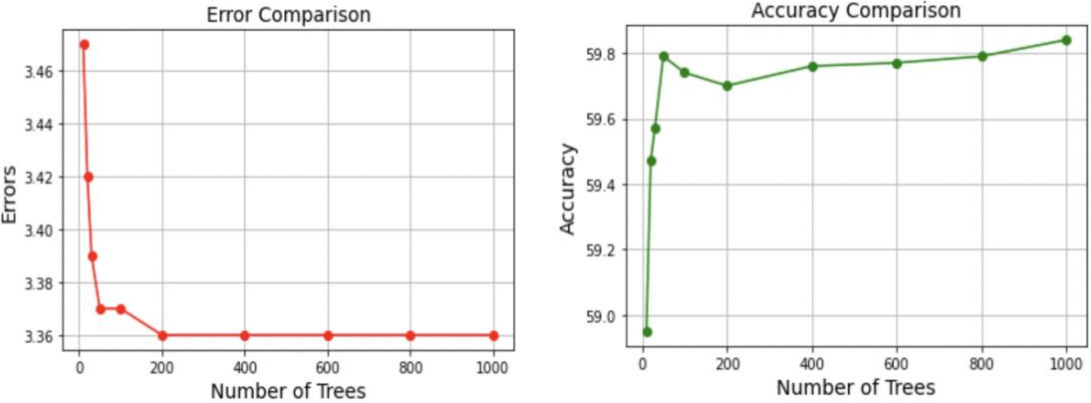

*Figure 5 Error and Accuracy Comparison for Random Forest Regression*

## Poisson and Negative Binomial Regressions

We initially tried a Poisson regression to predict accident counts based on our designated explanatory variables. Poisson regressions are designed to model count data; however, the Poisson regression makes the assumption that the mean equals the variance. After an initial investigation we found that our data was over dispersed and thus did not follow a Poisson distribution. When count based data does not meet the criteria for a Poisson distribution, the Negative Binomial Regression (NB) model is generally recommended. The NB regression is considered a variant of the Poisson regression that includes additional parameters to account for over-dispersion. To run NB regressions the data set was then split into 80% for training the model and 20% for testing. The aggregate number of crashes is the independent variable for our model. The dependent variables are identified as the 24 dummy variables for hour, 7 dummy variables for weekdays, 12 dummy variables for month, and the aggregate indicator variable for precipitation. We computed a root mean square error value of 4 and compared it to the max value of 30, and minimum value of 5 within our predicted dataset.

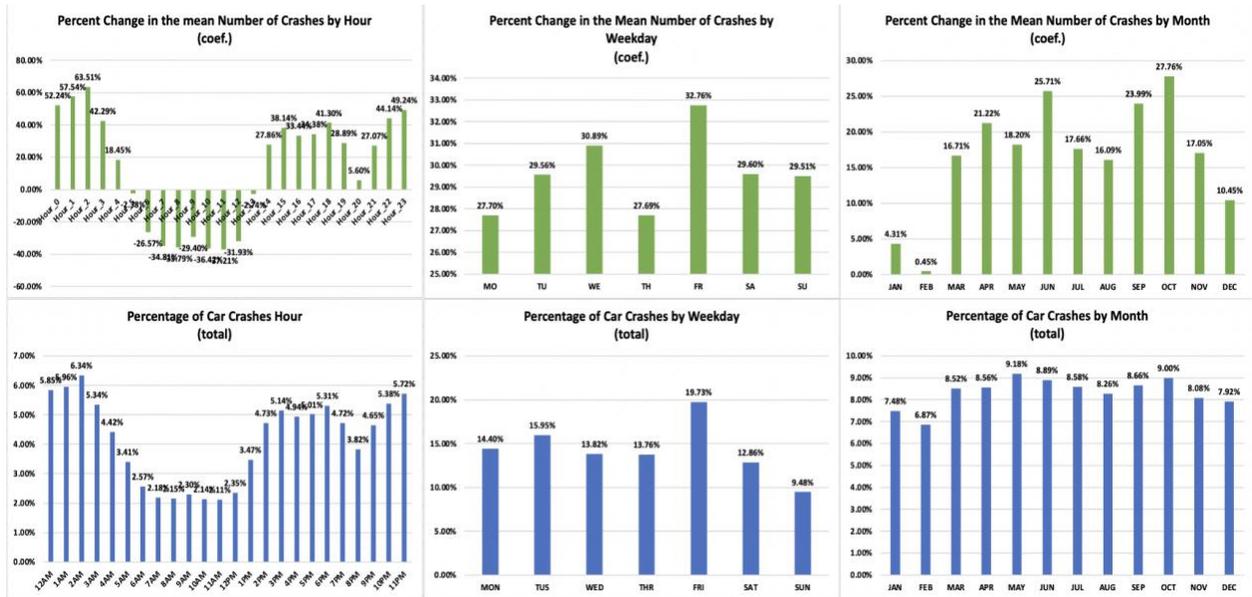

*Figure 6 Predicted impact each coefficient had on the crash count by Hour, Weekday and Month*

| Variables | Coefficent | exp(coef.) | Percent change | P-Score | Total Number of Crashes | Percentage of Total Number of Crashes |
|---|---|---|---|---|---|---|
| Hour_0 | 0.420 | 1.522 | 52.24% | 0.000 | 6222 | 5.85% |
| Hour_1 | 0.455 | 1.575 | 57.54% | 0.000 | 6338 | 5.96% |
| Hour_2 | 0.492 | 1.635 | 63.51% | 0.000 | 6743 | 6.34% |
| Hour_3 | 0.353 | 1.423 | 42.29% | 0.000 | 5681 | 5.34% |
| Hour_4 | 0.169 | 1.184 | 18.45% | 0.000 | 4701 | 4.42% |
| Hour_5 | -0.024 | 0.976 | -2.38% | 0.000 | 3627 | 3.41% |
| Hour_6 | -0.309 | 0.734 | -26.57% | 0.198 | 2733 | 2.57% |
| Hour_7 | -0.428 | 0.652 | -34.81% | 0.000 | 2320 | 2.18% |
| Hour_8 | -0.443 | 0.642 | -35.79% | 0.000 | 2293 | 2.15% |
| Hour_9 | -0.348 | 0.706 | -29.40% | 0.000 | 2446 | 2.30% |
| Hour_10 | -0.453 | 0.636 | -36.42% | 0.000 | 2276 | 2.14% |
| Hour_11 | -0.465 | 0.628 | -37.21% | 0.000 | 2250 | 2.11% |
| Hour_12 | -0.385 | 0.681 | -31.93% | 0.000 | 2504 | 2.35% |
| Hour_13 | -0.028 | 0.973 | -2.74% | 0.000 | 3694 | 3.47% |
| Hour_14 | 0.246 | 1.279 | 27.86% | 0.122 | 5030 | 4.73% |
| Hour_15 | 0.323 | 1.381 | 38.14% | 0.000 | 5469 | 5.14% |
| Hour_16 | 0.289 | 1.334 | 33.44% | 0.000 | 5259 | 4.94% |
| Hour_17 | 0.296 | 1.344 | 34.38% | 0.000 | 5336 | 5.01% |
| Hour_18 | 0.346 | 1.413 | 41.30% | 0.000 | 5651 | 5.31% |
| Hour_19 | 0.254 | 1.289 | 28.89% | 0.000 | 5020 | 4.72% |
| Hour_20 | 0.055 | 1.056 | 5.60% | 0.000 | 4069 | 3.82% |
| Hour_21 | 0.240 | 1.271 | 27.07% | 0.002 | 4950 | 4.65% |
| Hour_22 | 0.366 | 1.441 | 44.14% | 0.000 | 5723 | 5.38% |
| Hour_23 | 0.400 | 1.492 | 49.24% | 0.000 | 6087 | 5.72% |
| **Variables** | **Coefficient** | **exp(coef.)** | **Percent change** | **P-Score** | **Total Number of Crashes** | **Percentage of Total Number of Crashes** |
| Precipitation | 0.027 | 1.027 | 2.73% | 0.000 | | 0.00% |
| **Variables** | **Coefficient** | **exp(coef.)** | **Percent change** | **P-Score** | **Total Number of Crashes** | **Percentage of Total Number of Crashes** |
| MO | 0.245 | 1.277 | 27.70% | 0.000 | 15321 | 14.40% |
| TU | 0.259 | 1.296 | 29.56% | 0.000 | 16975 | 15.95% |
| WE | 0.269 | 1.309 | 30.89% | 0.000 | 14707 | 13.82% |
| TH | 0.244 | 1.277 | 27.69% | 0.000 | 14644 | 13.76% |
| FR | 0.283 | 1.328 | 32.76% | 0.000 | 20999 | 19.73% |
| SA | 0.259 | 1.296 | 29.60% | 0.000 | 13683 | 12.86% |
| SU | 0.259 | 1.295 | 29.51% | 0.000 | 10093 | 9.48% |
| **Variables** | **Coefficient** | **exp(coef.)** | **Percent change** | **P-Score** | **Total Number of Crashes** | **Percentage of Total Number of Crashes** |
| JAN | 0.042 | 1.043 | 4.31% | 0.000 | 7957 | 7.48% |
| FEB | 0.005 | 1.005 | 0.45% | 0.000 | 7311 | 6.87% |
| MAR | 0.155 | 1.167 | 16.71% | 0.720 | 9066 | 8.52% |
| APR | 0.192 | 1.212 | 21.22% | 0.000 | 9115 | 8.56% |
| MAY | 0.167 | 1.182 | 18.20% | 0.000 | 9774 | 9.18% |
| JUN | 0.229 | 1.257 | 25.71% | 0.000 | 9458 | 8.89% |
| JUL | 0.163 | 1.177 | 17.66% | 0.000 | 9133 | 8.58% |
| AUG | 0.149 | 1.161 | 16.09% | 0.000 | 8793 | 8.26% |
| SEP | 0.215 | 1.240 | 23.99% | 0.000 | 9214 | 8.66% |
| OCT | 0.245 | 1.278 | 27.76% | 0.000 | 9573 | 9.00% |
| NOV | 0.157 | 1.170 | 17.05% | 0.000 | 8598 | 8.08% |
| DEC | 0.099 | 1.105 | 10.45% | 0.000 | 8430 | 7.92% |

*Figure 7 Final coefficient for all variables*

The coefficients for our variables were exponentiated multiplicative factors by which the mean count changes. We then computed the percent change that our coefficients implied. The percentage of the coefficient tells us by what percent increase that variable would impact the mean crash count by 1 unit. The resulting matrix of all the variables can be found after the conclusion. Figure 6 are plots for each variable grouping (Hour, Weekday and Month) for the predicted impact each coefficient had on the mean crash count in our model, along with a corresponding graph of the percentage of the total number of crashes each variable represented in the whole dataset. The variables found to not be statistically significant based on a Wald-type test were Hours 6 and 14, and March (as shown in Figure 7). They were likely found to not be statistically significant because they were closely associated with the mean.

## Discussion and Conclusion

We analyzed 106,422 crashes over 4 years (2016-2019) from Washington, DC. Our project intended to find a relationship between time, crash data, and weather data. We created two models that predicted actual crashes to a reasonable degree of accuracy. We found that the least likely day to crash on was Friday, the least likely month was October, and the least likely hour was 2pm. Conversely, the most likely day was Thursday, the most likely month was February, and the most likely hour was 11am. The main application for our model we considered was as a tool to assist in commuting. We anticipated, with the increasing ability for people to telework, that a tool using this model could be used to determine days in which crashes are more likely to occur. Ideally, this would allow people to pick the most desirable days to choose not to commute, and even possibly have the effect of reducing the likelihood of crashes in the first place.

Future work building on our study would include incorporating traffic volume data in the models. We attempted to use data publicly available. However, it only had average annual daily traffic (AADT) for all areas of DC, rather than a count by date, making it challenging to find a sensible purpose to merge it with our other data. As a result, we decided not to use the volume data, but we recognize that it would be a useful addition if data was available in a more granular (at least weekly, if not daily) fashion. Granularity of weather data is another way to enhance the models. We only had access to weather data by day. With more granular weather data, our model could be improved to better match reality. Incorporation of precise GIS location data for crashes could potentially be integrated into the model we created to improve its efficacy. Additionally, it could be used in an implementation, to show potential "hotspots" where a crash is predicted more likely to occur, potentially allowing commuters to take other routes. As a longer-term focus, determining the type of areas where frequent hot spots for crashes occur could assist in future city planning, or redesigns of what already exists.